\documentclass[aps,floatfix,twocolumn,notitlepage,superscriptaddress,preprintnumbers,nofootinbib,10pt]{revtex4-2}

\usepackage{natbib,setspace}
\usepackage{graphicx,soul}
\usepackage{amssymb}
\usepackage{amsmath,amssymb}
\usepackage{color}
\usepackage{fancyhdr}
\definecolor{darkblue}{rgb}{0,0,0.5}
\usepackage[colorlinks,linkcolor=darkblue,citecolor=darkblue,urlcolor=darkblue]{hyperref}

\allowdisplaybreaks

\newcommand\beq{\begin{equation}}
\newcommand\eeq{\end{equation}}

\newcommand\hc{\text{h.c.}}

\def\CP{$CP$}
\def\BR{\text{BR}}

\begin{document}

\title{A New Higgs Boson with Electron-Muon Flavor-Violating Couplings}


\author{R.~Primulando}
\email{rprimulando@unpar.ac.id}
\affiliation{Center for Theoretical Physics, Department of Physics, Parahyangan Catholic University, Jalan Ciumbuleuit 94, Bandung 40141, Indonesia}

\author{J. Julio} 
\email{julio@brin.go.id}
\affiliation{National Research and Innovation Agency, KST B.\,J.\,Habibie, South Tangerang 15314, Indonesia}

\author{N. Srimanobhas}
\email{norraphat.s@chula.ac.th}
\affiliation{High Energy Physics Research Unit, Department of Physics, Faculty of Science, Chulalongkorn University, Pathumwan, Bangkok 10330, Thailand}

\author{P. Uttayarat}
\email{patipan@g.swu.ac.th}
\affiliation{Department of Physics, Srinakharinwirot University, 114 Sukhumvit 23 Rd., Wattana, Bangkok
10110, Thailand}

\begin{abstract}
A recent CMS search for a new resonance decaying to $e\mu$ in the mass range 110 GeV to 160 GeV finds an excess of events at 146 GeV. We interpret the search results in the context of the type-III two-Higgs-doublet-model. We find that the excess is moderately constrained by low-energy lepton-flavor-violation constraints, in particular the $\mu\to e \gamma$ decay. We also find the bounds from CMS search can be superior to the low-energy constraints for the scalar mass between 110 GeV and 150 GeV, suggesting the importance of this mass range for future searches.
\end{abstract}

\maketitle

\flushbottom
\section{Introduction}

As the most recently discovered particle of the Standard Model (SM)~\cite{ATLAS:2012yve,CMS:2012qbp}, the 125-GeV Higgs boson, $h$, is the least studied fundamental particle. After the discovery, immense amount of works on precision measurements of the $h$ properties has been done at the Large Hadron Collider (LHC). It is found that the $h$ properties agree quite well with the SM expectations~\cite{CMS:2022dwd,ATLAS:2022vkf}. In particular, since its discovery in the $\gamma\gamma$, $W^+W^-$ and $ZZ$ channels, the fermionic decay channels $b\bar b$~\cite{ATLAS:2018kot, CMS:2018nsn}, $\tau^+\tau^-$~\cite{ATLAS:2016neq}, and $\mu^+\mu^-$~\cite{CMS:2020xwi} have been established.  The consistency of these measurements with the SM predictions is one of the biggest triumphs of the SM. Searches for the remaining predicted decay channels, e.g., $e^+e^-$~\cite{CMS:2022urr,ATLAS:2019old} and $c\bar c$~\cite{ATLAS:2022ers,CMS:2022psv}, are underway. Despite all these, one should not discount other possible decay modes that are not predicted by the SM. Discovering any of these decays will be a clear signal of new physics. 

Lepton-flavor-violating (LFV) couplings are prototypical examples of new physics. The LFV couplings of the $h$ lead to LFV decays $h\to\ell\ell'$, which are correlated with low-energy LFV decays of charged lepton $\ell\to\ell'\gamma$ and $\ell \to 3 \ell'$. Additionally, the $\mu\to e$ conversion in atomic nuclei constrains any processes involving the LFV $e$-$\mu$ coupling. In the case of LFV tau sector, it has been demonstrated that the LHC searches for $h\to\tau\ell$ provide more stringent constraints than the low-energy processes~\cite{Blankenburg:2012ex,Harnik:2012pb}. This indicates that collider searches for $h\to\tau\ell$, together with their heavy scalar counterparts, are important in constraining new physics parameter space~\cite{Davidson:2010xv,Crivellin:2013wna,Kopp:2014rva,Buschmann:2016uzg,Primulando:2016eod,Altmannshofer:2016zrn,Primulando:2019ydt,Barman:2022iwj}.

The collider search for $h\to e\mu$, on the other hand, has often been overlooked as tools for constraining new physics parameter space. This is due to a conventional wisdom that such a channel is correlated with $\mu\to e\gamma$ and $\mu\to e$ conversion, which provide more stringent constraints~\cite{Blankenburg:2012ex,Harnik:2012pb}.  In this work, we show that such expectation does not generally apply to a new resonance decaying into $e^\pm\mu^\mp$. In particular, when the new particle mixes with the $h$ and its mass is less than 160 GeV, its contribution naturally cancels that of the $h$ in the $\mu\to e \gamma$ and $\mu\to e$ conversion, weakening the bounds from these processes.  

The strongest bound on $h\to e\mu$ derived from the full LHC Run II data is set by the CMS with $\text{BR}(h\to e^\pm\mu^\mp)\le 4.4\times10^{-5}$~\cite{CMS:2023pte}.  Beside searching for $h\to e^\pm\mu^\mp$, CMS also looks for the LFV decay of a new resonance, $H$, for 110 GeV $< m_H <$ 160 GeV. The CMS finds an excess at $m_H \sim 146$ GeV with a $3.8\sigma$ $(2.8\sigma)$ local (global) significance; the best fit cross-section is found to be $\sigma\left(pp \to H\to e^\pm\mu^\mp\right) = 3.82_{-1.09}^{+1.16}$ fb. However, the corresponding ATLAS search~\cite{ATLAS:2019old} does not find any significant excess at 146 GeV. Hence, the nature of the CMS excess remains inconclusive.

The CMS excess will inevitably induce $\mu\to e$ conversion in nuclei by a tree-level exchange of $H$. One can estimate the $\mu\to e$ conversion rate in a simplified model where $H$ is assumed to couple only to  $e$-$\mu$ and to gluon via an effective coupling. We find that in this scenario the induced rate for $\mu\to e$ conversion in the gold nucleus can be as low as $10^{-16}$, which is safely below current experimental bound.  However, in a more complete model, there will be more states that contribute to the process. Hence, it is compelling to investigate this in a more realistic model.

In this work, we compare the CMS search results against the corresponding low-energy bounds in the context of the type-III two-Higgs-doublet model (2HDM). While not advocating the presence of the excess, we try to examine whether it holds up against the current and the future $\mu\to e \gamma$ and $\mu \to e$ conversion bounds. We also highlight the importance of collider searches for LFV decays at a relatively low mass, i.e., between 100 GeV and 150 GeV.

\section{Type-III 2HDM}
\label{sec:model}
The type-III 2HDM is conveniently described in the Higgs basis~\cite{Georgi:1978ri}, where the two Higgs doublets, $H_{1,2}$, are given by
\begin{equation}
	H_1 = \begin{pmatrix}G^+\\[0.2em] \dfrac{v+h_1+iG}{\sqrt{2}} \end{pmatrix},\quad
	H_2 = \begin{pmatrix}H^+ \\[0.2em] \dfrac{h_2+iA}{\sqrt{2}}\end{pmatrix}.
	\label{eq:doublet}
\end{equation}
Here $v$ is the vacuum expectation value, and $G^+$ and $G$ are the would-be Goldstone bosons. For simplicity, we assume a \CP~symmetry in the scalar sector so that the two \CP-even states, $h_1$ and $h_2$, do not mix with the \CP-odd state $A$. Both $h_1$ and $h_2$ can mix through
\begin{equation}
	\begin{pmatrix}h_1\\h_2\end{pmatrix} = \begin{pmatrix}c_\alpha &s_\alpha\\-s_\alpha &c_\alpha\end{pmatrix} \begin{pmatrix}h\\H\end{pmatrix}, 
 \label{eq:mix}
\end{equation}
where $c_\alpha$ ($s_\alpha$) stands for $\cos\alpha$ ($\sin\alpha$). Here we identify $h$ with the 125-GeV Higgs boson. Since the properties of the $h$ agree with the SM predictions~\cite{Khachatryan:2016vau,CMS:2018uag,ATLAS:2019nkf}, the mixing angle $s_\alpha$ is expected to be small. The masses of the extra Higgs bosons $H$, $A$, and $H^+$ are, in principle, arbitrary. 

The Yukawa couplings of $H_1$ generate fermion masses while the Yukawa couplings of $H_2$ give rise to potential flavor violations. The most general form of the Yukawa couplings of $H_1$ and $H_2$ to the leptons is given by 
\begin{equation}
\begin{split}
	\mathcal{L}_{yuk} &\supset -\frac{\sqrt{2}m_{i}}{v}\bar{L}_i\ell_{Ri} H_1 - \sqrt{2}Y_{ij}\bar{L}_i\ell_{Rj}H_2 + \hc,
\end{split}
\label{eq:yukbeforeewsb}
\end{equation}
where $L$ denotes the lepton doublet, 
$m_i$ is the $i$th generation charged lepton mass, and $i,j=e,\mu,\tau$ indicate lepton generations.

In this work, we will focus on LFV in the  $e$-$\mu$ sector, so we take only $Y_{e\mu}$ and $Y_{\mu e}$ in Eq.~\eqref{eq:yukbeforeewsb} to be nonzero. The couplings $Y_{e\mu}$ and $Y_{\mu e}$, in principle, can be complex. However, the imaginary part of $Y_{e\mu}Y_{\mu e}$ is strongly constrained by the electron electric dipole moment measurement~\cite{ACME:2018yjb}.  Therefore, for simplicity, we assume that both $Y_{e\mu}$ and $Y_{\mu e}$ are real in this work.

In our minimal scenario, only $h$ and $H$ can be singly produced via gluon fusion and vector boson fusion processes. This makes $H$ the only relevant resonance for CMS and ATLAS LFV searches. The $H$ production cross-sections and non-LFV partial decay widths can be obtained from the would-be SM Higgs boson values by scaling them with a factor of $s_\alpha^2$.  On the other hand, the $h$ production cross-sections and non-LFV partial decay widths are reduced by a factor of $c_\alpha^2$ from their SM values. In our analysis, we use the SM-like Higgs cross-sections and decay widths provided by the LHC Higgs Cross Section Working Group~\cite{LHCHiggsCrossSectionWorkingGroup:2016ypw}.

\section{Low-energy LFV constraints}
\label{sec:lfv}
Naturally, the LFV couplings $Y_{e\mu}$ and $Y_{\mu e}$, together with the scalar mixing $s_\alpha$, will induce $\mu\to e\gamma$, which proceeds via loops. The partial decay width is given by
\begin{equation}
	\Gamma(\mu\to e\gamma) = \frac{\alpha_{em} m_\mu^5}{64\pi^4}\left(|c_L|^2 + |c_R|^2\right),
\end{equation}
where $\alpha_{em}$ is the electromagnetic fine-structure constant. The Wilson coefficients $c_{L,R}$ arise at one- and two-loop level. The one-loop contributions are given by
\begin{align}
   c_L^{(1)} = -\frac{s_{2\alpha}}{24}\frac{m_\mu Y_{\mu e}}{v}&\left[\frac{1}{m_h^2}\left(4 + 3\ln\frac{m_\mu^2}{m_h^2}\right)\right.\nonumber\\ 
   &\quad\left.- \frac{1}{m_H^2}\left(4 + 3\ln\frac{m_\mu^2}{m_H^2}\right) \right],
   \label{eq:1-loop}
\end{align}
with $s_{2\alpha}\equiv\sin 2 \alpha$. 
The main contributions to the Wilson coefficients $c_{L,R}$ are dominated by two-loop photon-exchange diagrams involving the $W$ boson and top quark. They are given by 
\begin{align}
    c_{L}^{(2W)} &\simeq -\frac{\alpha_{em} s_{2\alpha}Y_{\mu e}}{8\pi vm_\mu}\left[ \frac{f(z_{Wh}) - g(z_{Wh})}{2z_{Wh}} + 3f(z_{Wh}) \right.\nonumber\\ 
    &\quad\left. +\frac{23}{4}g(z_{Wh})+\frac{3}{4}h(z_{Wh}) - (h\to H)\right]\\
	c_L^{(2t)} &\simeq \frac{\alpha_{em} s_{2\alpha}Y_{\mu e}}{3\pi vm_\mu}\left[f(z_{th})- f(z_{tH}) \right],
        \label{eq:2-loop}
\end{align}
with $z_{ab} = m_a^2/m_b^2$. The loop functions $f$, $g$, and $h$ can be found in Ref.~\cite{Chang:1993kw}. Note the cancellation between the $h$ and the $H$ contributions due to the mixing of Eq.~\eqref{eq:mix}. Such a cancellation weakens the $\mu\to e \gamma$ constraint as $m_H$ approaches $m_h$. It should be noted that the loop functions in Eqs.~\eqref{eq:1-loop}--\eqref{eq:2-loop} are typically $\mathcal{O}(1)$. Hence, the one-loop coefficient is parametrically suppressed by $\,m_{\mu}^2/(\alpha_{em}m_{h(H)}^2)$ compared to the two-loop ones.
The Wilson coefficient $c_R$ is obtained from $c_L$ by replacing $Y_{\mu e}\to Y_{e\mu}$. 

In addition to such $W$ and top contributions, we have included contributions from the so-called ``set C'' diagrams~\cite{Chang:1993kw}. Numerically, they give a correction of order 10\% to the Wilson coefficients. Other contributions, i.e., 
two-loop $Z$-exchange diagrams are found to be more suppressed compared to the others.

The most stringent constraint on the $\mu\to e\gamma$ decay is provided by the MEG experiment with $\text{BR}(\mu\to e\gamma)\le4.2\times10^{-13}$~\cite{MEG:2016leq}. The upgraded MEGII experiment, which is currently taking data, is expected to push the bound down to $6\times10^{-14}$~\cite{MEGII:2018kmf}.

The Wilson coefficients $c_L$ and $c_R$ also lead to $\mu\to e$ conversion in atomic nuclei. In addition, the $\mu\to e$ conversion also gets tree-level contributions mediated by the \CP-even Higgs bosons. In our scenario, the conversion rate is given by~\cite{Kitano:2002mt} (see also \cite{Harnik:2012pb})
\begin{equation}
	\Gamma(\mu\to e) = m_\mu^5\left|\frac{e\,c_R}{16\pi^2}D + \sum_{N=p,n} g_{L}^NS^N\right|^2 + (L\leftrightarrow R),
\end{equation}
where $D$ and $S^N$ are overlap integrals, whose values are given in \cite{Kitano:2002mt}.  
The effective coupling $g_L^N$ is given by
\begin{equation}
	g_L^N = s_{2\alpha}Y_{e\mu}\left(\frac{1}{m_H^2}-\frac{1}{m_h^2}\right)\frac{m_N}{v}\sum_qf^{(q,N)},
\end{equation}
where $f^{(q,N)}$ and $m_N$ denote the nucleon form factor and the nucleon mass, respectively. The form factors are given in Refs.~\cite{Crivellin:2014cta,Bishara:2015cha}, and in above equation, they are summed over all quark flavors. The effective coupling $g_R^N$ can be obtained from $g_L^N$ by replacing $Y_{e\mu}\to Y_{\mu e}$. As in the $\mu\to e\gamma$ case, the $\mu\to e$ conversion constraint also suffers from the same blind spot when $m_H$ approaches $m_h$. Currently, the strongest constraint comes from conversion in the gold nucleus, which sets $\Gamma(\mu\to e$)/$\Gamma$(captured) $< 7\times10^{-13}$~\cite{Bertl:2006up}. 

\section{Collider Searches for LFV decays}
\label{sec:collider}

The LFV couplings $Y_{e\mu}$ and $Y_{\mu e}$, together with the mixing angle $s_\alpha$, lead to LFV decays of the $h$ and $H$. The partial decay width $h\to e^\pm\mu^\mp$ is given by
\begin{equation}
    \Gamma(h\to e^\pm\mu^\mp) = \frac{s_\alpha^2 m_h}{8\pi}\left(\left|Y_{e\mu}\right|^2 + \left|Y_{\mu e}\right|^2\right).
    \label{eq:LFVh}
\end{equation}
The partial decay width for $H\to e^\pm\mu^\mp$ can be obtained from Eq.~\eqref{eq:LFVh} by changing $s_\alpha \to c_\alpha$ and $m_h\to m_H$. 

Both ATLAS and CMS Collaborations have searched for such LFV decay of $h$ without any positive signal~\cite{CMS:2023pqk,ATLAS:2019old}. Assuming the SM production cross-sections for the $h$, the ATLAS has set an upper bound $\text{BR}({h\to e^\pm\mu^\mp})\le6.2\times10^{-5}$, while the CMS has set a slightly stronger constraint $\text{BR}({h\to e^\pm\mu^\mp})\le4.4\times10^{-5}$. As mentioned earlier, in our scenario, the $h$ production cross-sections get corrected by $c_\alpha^2$, so does the bound on the $h\to e^\pm\mu^\mp$ branching ratio. Hence, the CMS upper bound now reads $c_\alpha^2 \text{BR}({h\to e^\pm\mu^\mp})\le4.4\times10^{-5}$.

The collider search for $H\to e^\pm\mu^\mp$ was done by the CMS Collaboration~\cite{CMS:2023pqk}. The result was presented in terms of the $H$ production cross-section times the branching ratio into $e^\pm\mu^\mp$. In our model, it is formulated as
\begin{equation}
    \sigma\times \BR({H\to e^\pm\mu^\mp}) = \frac{s_\alpha^2\sigma_{SM}\Gamma(H\to e^\pm\mu^\mp)}{s_\alpha^2\Gamma_{SM} + \Gamma(H\to e^\pm\mu^\mp)},
\end{equation}
where $\sigma_{SM}$ and $\Gamma_{SM}$ are the cross-section and the total decay width of the SM-like Higgs with mass $m_H$. 

\subsection{The CMS 146-GeV excess}

The CMS search for a resonance decaying to $e^\pm\mu^\mp$ has observed an excess of events around 146 GeV with a local significance of 3.8$\sigma$~\cite{CMS:2023pqk}. However, the corresponding ATLAS search does not find any excess at 146 GeV. By performing a simple profiled likelihood ratio analysis, we find that the CMS signal is disfavored by the ATLAS measurement at the 2.5$\sigma$ level.

Taking both CMS and ATLAS results into consideration, we perform a naive combination of the two searches using the counting experiment. The combined tool with asymptotic approximation~\cite{Cowan_2011} is used to calculate the best fit cross-section and its significance. The data and background Monte Carlo (MC) from both experiments, as well as the CMS signal MC for $m_H = 146$ GeV, are extracted from the Data-MC plots in Refs.~\cite{CMS:2023pqk,ATLAS:2019old}. In the case of ATLAS MC, we estimate the signal model using Madgraph5~\cite{Alwall:2014hca} to generate parton-level productions and $H$ decays, followed by showering and hadronization with Pythia8~\cite{Bierlich:2022pfr}. Finally, the detector responses are simulated using Delphes3~\cite{deFavereau:2013fsa}. The overall uncertainty is estimated to be 1\%, as mentioned in Ref.~\cite{ATLAS:2019old} for the $e\mu$ channel. We find that the combination reduces the local significance of the excess to 3.3$\sigma$, with the best fit cross section $\sigma\times \BR({H\to e^\pm\mu^\mp}) = 2.92^{+0.91}_{-0.89}$ fb. 

\begin{figure}[ht]
\centering \includegraphics[width=8.6 cm]{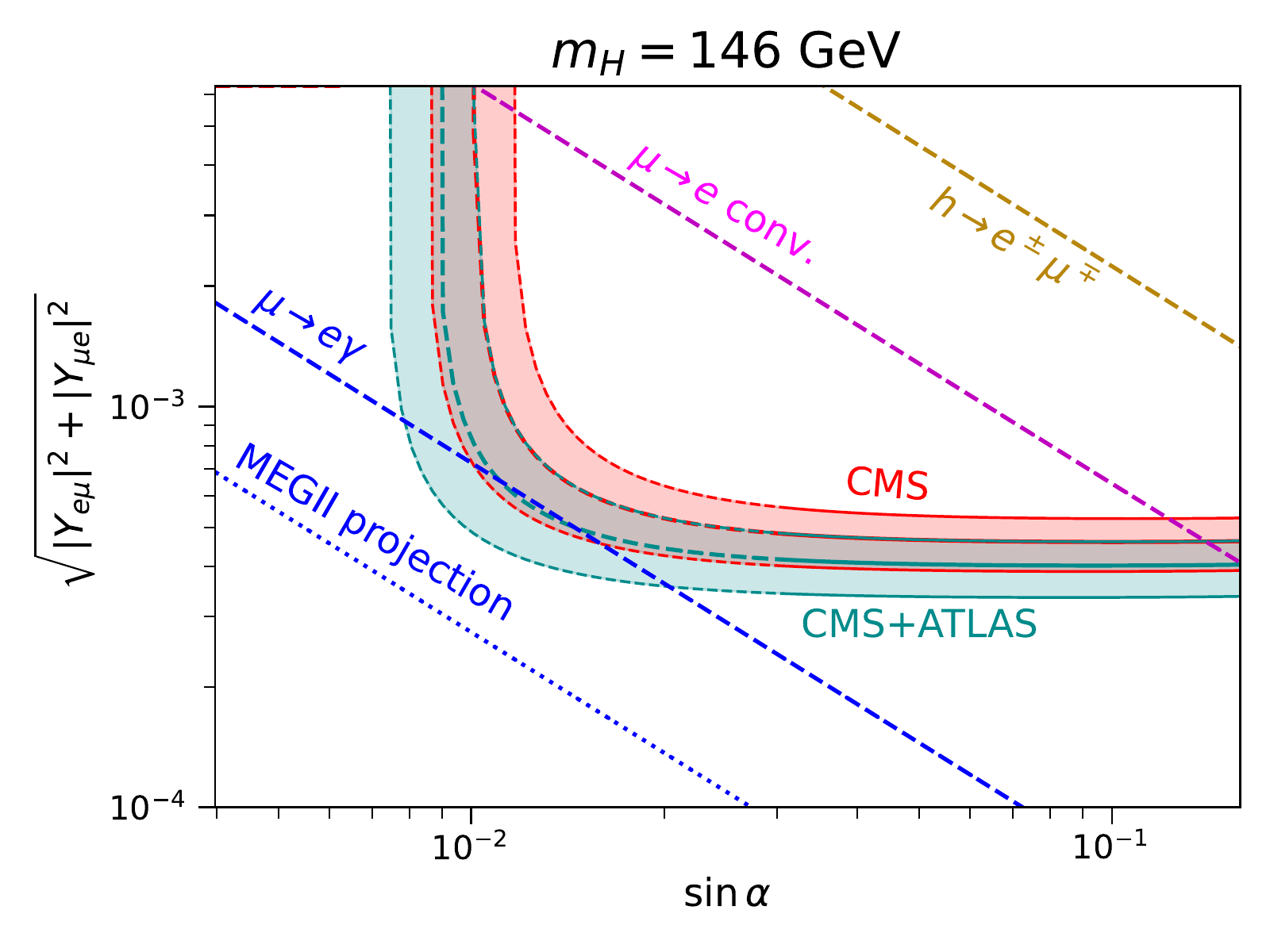}
\caption{The region in the $\sin\alpha$--$\sqrt{\left|Y_{e\mu}\right|^2+\left|Y_{\mu e}\right|^2}$ plane compatible with the CMS excess (red) and the CMS-ATLAS combination (cyan) at the 1$\sigma$ level. The region surrounded by the red (cyan) dashed lines is excluded by CMS multilepton search unless $m_A \gtrsim 800$ GeV. The yellow, magenta, and blue lines are the upper bounds from the $h\rightarrow e^\pm\mu^\mp$, $\mu\to e$ conversion, and  $\mu \rightarrow e \gamma$ searches, respectively.}
\label{fig:146comb}
\end{figure}
The region in $s_\alpha$--$\sqrt{Y_{e\mu}^2+Y_{\mu e}^2}$ parameter space consistent with the CMS excess, together with the combined CMS and ATLAS data, at 1$\sigma$ level is shown in Fig.~\ref{fig:146comb}. The preferred regions of parameter space are compared against constraints from the $h\to e\mu$, $\mu\to e\gamma$, and $\mu\to e$ conversion searches.  Note that the small 1$\sigma$ region preferred by the CMS excess consistent with the $\mu\to e\gamma$ bound is centered around $s_\alpha = 0.014$ and $\sqrt{Y_{e\mu}^2+Y_{\mu e}^2} =6.6\times10^{-4}$. In that figure, we also show the projected sensitivity of the MEGII experiment. Should the origin of the CMS excess be a new particle, the MEGII experiment will observe the $\mu\to e\gamma$ decay.

In addition to the aforementioned constraints, the 146-GeV excess is also constrained by the CMS search for a new resonance decaying into a pair of $ZZ^*$~\cite{CMS:2018amk} and a pair of tau leptons~\cite{CMS:2022goy}. In both cases, the bounds are $\mathcal{O}(100)$ fb, which correspond to $s_\alpha\lesssim0.2$.

The 146-GeV excess is also constrained by multilepton searches via $HA$ pair production. The $A$ will decay predominantly into $e\mu$, $hZ$, or $HZ$ if the latter is kinematically open.  A significant portion of the $H$, on the other hand, will decay into $e\mu$ for sufficiently small $s_\alpha$. If this is the case, the pair produced $HA$ will result in four-lepton signatures, which are strongly constrained by the CMS multilepton searches~\cite{CMS:2021cox}.  
From our analysis, we find that the CMS multilepton constraints imply either  $m_A\gtrsim800$ GeV or  $s_\alpha\gtrsim0.025$ provided that $A\to HZ$ is kinematically open.

\subsection{Bounds for other masses}
While  the excess at 146 GeV seen by CMS could be a result of a new particle, we should not draw a definite conclusion on its nature without more data. Thus, we also study the possibility that the CMS search finds no excess anywhere in the search region 110 GeV $\le m_H\le$ 160 GeV. In this scenario, we take the 95\% confidence level upper limits on $\sigma\times \BR({H\to e\mu})$ reported by CMS and compare them against constraints from $\mu\to e\gamma$ and $\mu\to e$ conversion.     

\begin{figure}[t]
\includegraphics[width=8.6 cm]{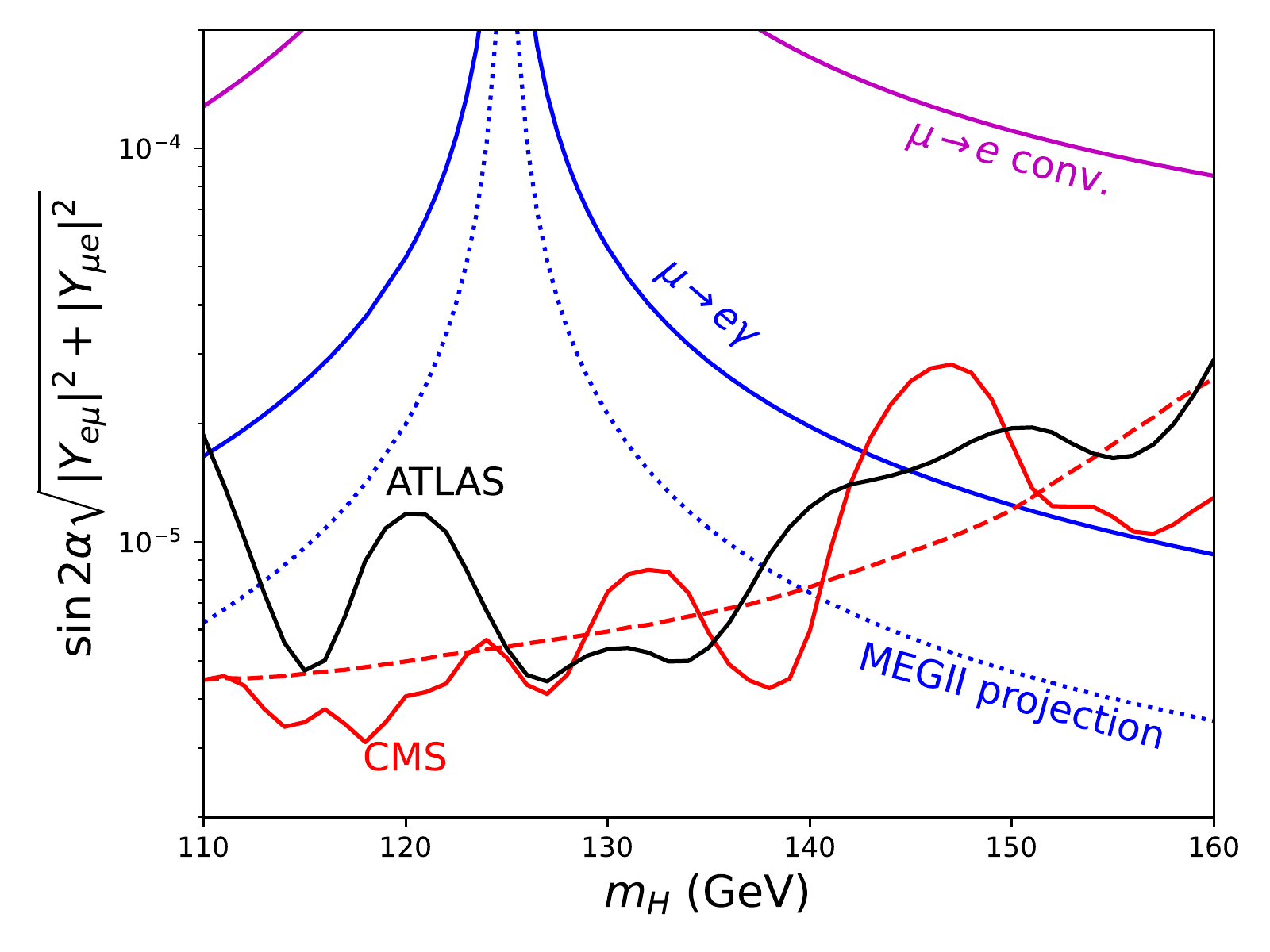}
\caption{The comparison between CMS observed limit (solid red), CMS expected limit (dashed red), ATLAS observed limit (black), current $\mu\to e\gamma$ constraint (solid blue), current $\mu\to e$ conversion (magenta), and the projected $\mu\to e\gamma$ limit from MEGII (dotted blue) on the CMS search region. The bounds are projected onto the $m_H-s_{2\alpha}\sqrt{\left|Y_{e\mu}\right|^2+\left|Y_{\mu e}\right|^2}$ plane. For the CMS bounds, only the lower end of the projection is shown.}
\label{fig:boundall}
\end{figure}

The bounds from collider search, $\mu\to e\gamma$, and $\mu\to e$ conversion experiments are functions of the mixing angle $s_\alpha$ and the LFV couplings $\sqrt{\left|Y_{e\mu}\right|^2+\left|Y_{\mu e}\right|^2}$. While the $\mu\to e\gamma$ and $\mu\to e$ conversion constraints depend on the product $s_{2\alpha}\sqrt{\left|Y_{e\mu}\right|^2+\left|Y_{\mu e}\right|^2}$,  the collider constraint is a more complicated function. For illustrative purposes, the collider bound, for certain value of $H$ mass, would trace out a curve in the $s_\alpha$--$\sqrt{\left|Y_{e\mu}\right|^2+\left|Y_{\mu e}\right|^2}$ plane similar to the preferred region for the 146-GeV excess in Fig.~\ref{fig:146comb}.

In order to compare these two types of constraints on the entire CMS search region, we project the constraints onto the $s_{2\alpha}\sqrt{\left|Y_{e\mu}\right|^2+\left|Y_{\mu e}\right|^2}$  line. Such a projection maps the $\mu\to e\gamma$ and $\mu\to e$ conversion constraints to a point. On the other hand, the bound from CMS search gets mapped into a bounded-from-below interval. If the minimum of such an interval is smaller than the $\mu\to e\gamma$ ($\mu\to e$ conversion) projection, it is said that the collider bound is more constraining than the $\mu\to e\gamma$  ($\mu\to e$ conversion) bound for some part of the $s_\alpha$--$\sqrt{\left|Y_{e\mu}\right|^2+\left|Y_{\mu e}\right|^2}$ parameter space. In Fig.~\ref{fig:boundall} we show the comparison between the collider bounds and the $\mu\to e\gamma$ and $\mu\to e$ conversion constraints for 110 GeV $\le m_H\le$160 GeV. Note that, in Fig.~\ref{fig:boundall}, only the minimum of the collider constraint projection is shown for each $m_H$ value. As can be seen from the plot, for $m_H\lesssim$ 140 GeV, the collider search can be more constraining than even the projected MEGII bound. Hence, we advocate both the ATLAS and CMS Collaborations to continue searching for $H \rightarrow e^\pm\mu^\mp$ in that low-$m_H$ region.

\section{Conclusion and Discussion}
\label{sec:conc}
In this work, we have analyzed the LHC searches for a new resonance decaying to $e^\pm\mu^\mp$ in the context of the type-III 2HDM. 
This analysis is motivated by a recent CMS search~\cite{CMS:2023pqk} that explores the LFV decay of a new scalar boson between 110 GeV $< m_H < 160$ GeV. Within the search region, the CMS finds a possible excess at $m_H = 146$ GeV with a local (global) significance of $3.8\sigma$ $(2.8\sigma)$. A simplistic combination of the CMS and ATLAS searches reduces the local significance to 3.3$\sigma$, with  $\sigma\times \BR({H\to e^\pm\mu^\mp}) = 2.92^{+0.91}_{-0.89}$ fb. In the type-III 2HDM context, the 146-GeV excess is only moderately constrained by the current $\mu\to e \gamma$ data, while the future MEGII search will probe the whole parameter region preferred by the excess. 

It is interesting to note that the 146-GeV excess could be related to another excess from the so-called ``multilepton anomalies''~\cite{Crivellin:2021ubm}. We leave this investigation for possible future work.

In the event that the excess 
is due to an upward fluctuation in the data, we analyze the bounds on $\sigma\times\text{BR}({H\to e^\pm\mu^\mp})$ provided by CMS over the whole search region. 
The comparison between the CMS bounds and the low-energy counterparts is shown in Fig.~\ref{fig:boundall}. From the plot, we see that for 110 GeV $\le m_H \lesssim$ 140 GeV, the current CMS bound is better than the current and even the projected MEGII bounds for $\mu\to e \gamma$.
Thus, we encourage both CMS and ATLAS Collaborations to continue searching for LFV decays of a new resonance in this low-mass region.   

Lastly, we note that even though we take $Y_{ee}$ and $Y_{\mu\mu}$ to be zero in this work, it does not mean we need a large hierarchy between the flavor-violating and the flavor-conserving couplings. In fact, $Y_{ee/\mu\mu}$ can be comparable to $Y_{e\mu/\mu e}$ and still be consistent with low-energy constraints. We set them to zero in the spirit of minimal scenario.

\begin{acknowledgments}
R.\,P. was supported by the Parahyangan Catholic University under grant no. 
III/LPPM/2023-02/32-P. J.\,J. was supported in part by the Indonesia Toray Science Foundation.
P.\,U. was supported in part by the Mid-Career Research Grant from the National Research Council of Thailand under contract no.~N42A650378. N.\,S. was supported by Thailand NSRF via PMU-B under grant number B37G660013. R.\,P. thanks The Abdus Salam International Centre for Theoretical Physics for kind hospitality while this work was being completed.

\end{acknowledgments}

\bibliography{reference} 

\begin{thebibliography}{45}%
\makeatletter
\providecommand \@ifxundefined [1]{%
 \@ifx{#1\undefined}
}%
\providecommand \@ifnum [1]{%
 \ifnum #1\expandafter \@firstoftwo
 \else \expandafter \@secondoftwo
 \fi
}%
\providecommand \@ifx [1]{%
 \ifx #1\expandafter \@firstoftwo
 \else \expandafter \@secondoftwo
 \fi
}%
\providecommand \natexlab [1]{#1}%
\providecommand \enquote  [1]{``#1''}%
\providecommand \bibnamefont  [1]{#1}%
\providecommand \bibfnamefont [1]{#1}%
\providecommand \citenamefont [1]{#1}%
\providecommand \href@noop [0]{\@secondoftwo}%
\providecommand \href [0]{\begingroup \@sanitize@url \@href}%
\providecommand \@href[1]{\@@startlink{#1}\@@href}%
\providecommand \@@href[1]{\endgroup#1\@@endlink}%
\providecommand \@sanitize@url [0]{\catcode `\\12\catcode `\$12\catcode
  `\&12\catcode `\#12\catcode `\^12\catcode `\_12\catcode `\%12\relax}%
\providecommand \@@startlink[1]{}%
\providecommand \@@endlink[0]{}%
\providecommand \url  [0]{\begingroup\@sanitize@url \@url }%
\providecommand \@url [1]{\endgroup\@href {#1}{\urlprefix }}%
\providecommand \urlprefix  [0]{URL }%
\providecommand \Eprint [0]{\href }%
\providecommand \doibase [0]{http://dx.doi.org/}%
\providecommand \selectlanguage [0]{\@gobble}%
\providecommand \bibinfo  [0]{\@secondoftwo}%
\providecommand \bibfield  [0]{\@secondoftwo}%
\providecommand \translation [1]{[#1]}%
\providecommand \BibitemOpen [0]{}%
\providecommand \bibitemStop [0]{}%
\providecommand \bibitemNoStop [0]{.\EOS\space}%
\providecommand \EOS [0]{\spacefactor3000\relax}%
\providecommand \BibitemShut  [1]{\csname bibitem#1\endcsname}%
\let\auto@bib@innerbib\@empty
\bibitem [{\citenamefont {Aad}\ \emph {et~al.}(2012)\citenamefont {Aad} \emph
  {et~al.}}]{ATLAS:2012yve}%
  \BibitemOpen
  \bibfield  {author} {\bibinfo {author} {\bibfnamefont {G.}~\bibnamefont
  {Aad}} \emph {et~al.} (\bibinfo {collaboration} {ATLAS}),\ }\href {\doibase
  10.1016/j.physletb.2012.08.020} {\bibfield  {journal} {\bibinfo  {journal}
  {Phys. Lett. B}\ }\textbf {\bibinfo {volume} {716}},\ \bibinfo {pages} {1}
  (\bibinfo {year} {2012})},\ \Eprint {http://arxiv.org/abs/1207.7214}
  {arXiv:1207.7214 [hep-ex]} \BibitemShut {NoStop}%
\bibitem [{\citenamefont {Chatrchyan}\ \emph {et~al.}(2012)\citenamefont
  {Chatrchyan} \emph {et~al.}}]{CMS:2012qbp}%
  \BibitemOpen
  \bibfield  {author} {\bibinfo {author} {\bibfnamefont {S.}~\bibnamefont
  {Chatrchyan}} \emph {et~al.} (\bibinfo {collaboration} {CMS}),\ }\href
  {\doibase 10.1016/j.physletb.2012.08.021} {\bibfield  {journal} {\bibinfo
  {journal} {Phys. Lett. B}\ }\textbf {\bibinfo {volume} {716}},\ \bibinfo
  {pages} {30} (\bibinfo {year} {2012})},\ \Eprint
  {http://arxiv.org/abs/1207.7235} {arXiv:1207.7235 [hep-ex]} \BibitemShut
  {NoStop}%
\bibitem [{\citenamefont {Tumasyan}\ \emph
  {et~al.}(2022{\natexlab{a}})\citenamefont {Tumasyan} \emph
  {et~al.}}]{CMS:2022dwd}%
  \BibitemOpen
  \bibfield  {author} {\bibinfo {author} {\bibfnamefont {A.}~\bibnamefont
  {Tumasyan}} \emph {et~al.} (\bibinfo {collaboration} {CMS}),\ }\href
  {\doibase 10.1038/s41586-022-04892-x} {\bibfield  {journal} {\bibinfo
  {journal} {Nature}\ }\textbf {\bibinfo {volume} {607}},\ \bibinfo {pages}
  {60} (\bibinfo {year} {2022}{\natexlab{a}})},\ \Eprint
  {http://arxiv.org/abs/2207.00043} {arXiv:2207.00043 [hep-ex]} \BibitemShut
  {NoStop}%
\bibitem [{\citenamefont {{ATLAS Collaboration}}(2022)}]{ATLAS:2022vkf}%
  \BibitemOpen
  \bibfield  {author} {\bibinfo {author} {\bibnamefont {{ATLAS
  Collaboration}}},\ }\href {\doibase 10.1038/s41586-022-04893-w} {\bibfield
  {journal} {\bibinfo  {journal} {Nature}\ }\textbf {\bibinfo {volume} {607}},\
  \bibinfo {pages} {52} (\bibinfo {year} {2022})},\ \bibinfo {note} {[Erratum:
  Nature 612, E24 (2022)]},\ \Eprint {http://arxiv.org/abs/2207.00092}
  {arXiv:2207.00092 [hep-ex]} \BibitemShut {NoStop}%
\bibitem [{\citenamefont {Aaboud}\ \emph {et~al.}(2018)\citenamefont {Aaboud}
  \emph {et~al.}}]{ATLAS:2018kot}%
  \BibitemOpen
  \bibfield  {author} {\bibinfo {author} {\bibfnamefont {M.}~\bibnamefont
  {Aaboud}} \emph {et~al.} (\bibinfo {collaboration} {ATLAS}),\ }\href
  {\doibase 10.1016/j.physletb.2018.09.013} {\bibfield  {journal} {\bibinfo
  {journal} {Phys. Lett. B}\ }\textbf {\bibinfo {volume} {786}},\ \bibinfo
  {pages} {59} (\bibinfo {year} {2018})},\ \Eprint
  {http://arxiv.org/abs/1808.08238} {arXiv:1808.08238 [hep-ex]} \BibitemShut
  {NoStop}%
\bibitem [{\citenamefont {Sirunyan}\ \emph
  {et~al.}(2018{\natexlab{a}})\citenamefont {Sirunyan} \emph
  {et~al.}}]{CMS:2018nsn}%
  \BibitemOpen
  \bibfield  {author} {\bibinfo {author} {\bibfnamefont {A.~M.}\ \bibnamefont
  {Sirunyan}} \emph {et~al.} (\bibinfo {collaboration} {CMS}),\ }\href
  {\doibase 10.1103/PhysRevLett.121.121801} {\bibfield  {journal} {\bibinfo
  {journal} {Phys. Rev. Lett.}\ }\textbf {\bibinfo {volume} {121}},\ \bibinfo
  {pages} {121801} (\bibinfo {year} {2018}{\natexlab{a}})},\ \Eprint
  {http://arxiv.org/abs/1808.08242} {arXiv:1808.08242 [hep-ex]} \BibitemShut
  {NoStop}%
\bibitem [{\citenamefont {Aad}\ \emph {et~al.}(2016{\natexlab{a}})\citenamefont
  {Aad} \emph {et~al.}}]{ATLAS:2016neq}%
  \BibitemOpen
  \bibfield  {author} {\bibinfo {author} {\bibfnamefont {G.}~\bibnamefont
  {Aad}} \emph {et~al.} (\bibinfo {collaboration} {ATLAS, CMS}),\ }\href
  {\doibase 10.1007/JHEP08(2016)045} {\bibfield  {journal} {\bibinfo  {journal}
  {JHEP}\ }\textbf {\bibinfo {volume} {08}},\ \bibinfo {pages} {045} (\bibinfo
  {year} {2016}{\natexlab{a}})},\ \Eprint {http://arxiv.org/abs/1606.02266}
  {arXiv:1606.02266 [hep-ex]} \BibitemShut {NoStop}%
\bibitem [{\citenamefont {Sirunyan}\ \emph {et~al.}(2021)\citenamefont
  {Sirunyan} \emph {et~al.}}]{CMS:2020xwi}%
  \BibitemOpen
  \bibfield  {author} {\bibinfo {author} {\bibfnamefont {A.~M.}\ \bibnamefont
  {Sirunyan}} \emph {et~al.} (\bibinfo {collaboration} {CMS}),\ }\href
  {\doibase 10.1007/JHEP01(2021)148} {\bibfield  {journal} {\bibinfo  {journal}
  {JHEP}\ }\textbf {\bibinfo {volume} {01}},\ \bibinfo {pages} {148} (\bibinfo
  {year} {2021})},\ \Eprint {http://arxiv.org/abs/2009.04363} {arXiv:2009.04363
  [hep-ex]} \BibitemShut {NoStop}%
\bibitem [{\citenamefont {{CMS
  Collaboration}}(2022{\natexlab{a}})}]{CMS:2022urr}%
  \BibitemOpen
  \bibfield  {author} {\bibinfo {author} {\bibnamefont {{CMS Collaboration}}},\
  }\href@noop {} {\  (\bibinfo {year} {2022}{\natexlab{a}})},\ \Eprint
  {http://arxiv.org/abs/2208.00265} {arXiv:2208.00265 [hep-ex]} \BibitemShut
  {NoStop}%
\bibitem [{\citenamefont {Aad}\ \emph {et~al.}(2020{\natexlab{a}})\citenamefont
  {Aad} \emph {et~al.}}]{ATLAS:2019old}%
  \BibitemOpen
  \bibfield  {author} {\bibinfo {author} {\bibfnamefont {G.}~\bibnamefont
  {Aad}} \emph {et~al.} (\bibinfo {collaboration} {ATLAS}),\ }\href {\doibase
  10.1016/j.physletb.2019.135148} {\bibfield  {journal} {\bibinfo  {journal}
  {Phys. Lett. B}\ }\textbf {\bibinfo {volume} {801}},\ \bibinfo {pages}
  {135148} (\bibinfo {year} {2020}{\natexlab{a}})},\ \Eprint
  {http://arxiv.org/abs/1909.10235} {arXiv:1909.10235 [hep-ex]} \BibitemShut
  {NoStop}%
\bibitem [{\citenamefont {Aad}\ \emph {et~al.}(2022)\citenamefont {Aad} \emph
  {et~al.}}]{ATLAS:2022ers}%
  \BibitemOpen
  \bibfield  {author} {\bibinfo {author} {\bibfnamefont {G.}~\bibnamefont
  {Aad}} \emph {et~al.} (\bibinfo {collaboration} {ATLAS}),\ }\href {\doibase
  10.1140/epjc/s10052-022-10588-3} {\bibfield  {journal} {\bibinfo  {journal}
  {Eur. Phys. J. C}\ }\textbf {\bibinfo {volume} {82}},\ \bibinfo {pages} {717}
  (\bibinfo {year} {2022})},\ \Eprint {http://arxiv.org/abs/2201.11428}
  {arXiv:2201.11428 [hep-ex]} \BibitemShut {NoStop}%
\bibitem [{\citenamefont {{CMS
  Collaboration}}(2022{\natexlab{b}})}]{CMS:2022psv}%
  \BibitemOpen
  \bibfield  {author} {\bibinfo {author} {\bibnamefont {{CMS Collaboration}}},\
  }\href@noop {} {\  (\bibinfo {year} {2022}{\natexlab{b}})},\ \Eprint
  {http://arxiv.org/abs/2205.05550} {arXiv:2205.05550 [hep-ex]} \BibitemShut
  {NoStop}%
\bibitem [{\citenamefont {Blankenburg}\ \emph {et~al.}(2012)\citenamefont
  {Blankenburg}, \citenamefont {Ellis},\ and\ \citenamefont
  {Isidori}}]{Blankenburg:2012ex}%
  \BibitemOpen
  \bibfield  {author} {\bibinfo {author} {\bibfnamefont {G.}~\bibnamefont
  {Blankenburg}}, \bibinfo {author} {\bibfnamefont {J.}~\bibnamefont {Ellis}},
  \ and\ \bibinfo {author} {\bibfnamefont {G.}~\bibnamefont {Isidori}},\ }\href
  {\doibase 10.1016/j.physletb.2012.05.007} {\bibfield  {journal} {\bibinfo
  {journal} {Phys. Lett. B}\ }\textbf {\bibinfo {volume} {712}},\ \bibinfo
  {pages} {386} (\bibinfo {year} {2012})},\ \Eprint
  {http://arxiv.org/abs/1202.5704} {arXiv:1202.5704 [hep-ph]} \BibitemShut
  {NoStop}%
\bibitem [{\citenamefont {Harnik}\ \emph {et~al.}(2013)\citenamefont {Harnik},
  \citenamefont {Kopp},\ and\ \citenamefont {Zupan}}]{Harnik:2012pb}%
  \BibitemOpen
  \bibfield  {author} {\bibinfo {author} {\bibfnamefont {R.}~\bibnamefont
  {Harnik}}, \bibinfo {author} {\bibfnamefont {J.}~\bibnamefont {Kopp}}, \ and\
  \bibinfo {author} {\bibfnamefont {J.}~\bibnamefont {Zupan}},\ }\href
  {\doibase 10.1007/JHEP03(2013)026} {\bibfield  {journal} {\bibinfo  {journal}
  {JHEP}\ }\textbf {\bibinfo {volume} {03}},\ \bibinfo {pages} {026} (\bibinfo
  {year} {2013})},\ \Eprint {http://arxiv.org/abs/1209.1397} {arXiv:1209.1397
  [hep-ph]} \BibitemShut {NoStop}%
\bibitem [{\citenamefont {Davidson}\ and\ \citenamefont
  {Grenier}(2010)}]{Davidson:2010xv}%
  \BibitemOpen
  \bibfield  {author} {\bibinfo {author} {\bibfnamefont {S.}~\bibnamefont
  {Davidson}}\ and\ \bibinfo {author} {\bibfnamefont {G.~J.}\ \bibnamefont
  {Grenier}},\ }\href {\doibase 10.1103/PhysRevD.81.095016} {\bibfield
  {journal} {\bibinfo  {journal} {Phys. Rev. D}\ }\textbf {\bibinfo {volume}
  {81}},\ \bibinfo {pages} {095016} (\bibinfo {year} {2010})},\ \Eprint
  {http://arxiv.org/abs/1001.0434} {arXiv:1001.0434 [hep-ph]} \BibitemShut
  {NoStop}%
\bibitem [{\citenamefont {Crivellin}\ \emph {et~al.}(2013)\citenamefont
  {Crivellin}, \citenamefont {Kokulu},\ and\ \citenamefont
  {Greub}}]{Crivellin:2013wna}%
  \BibitemOpen
  \bibfield  {author} {\bibinfo {author} {\bibfnamefont {A.}~\bibnamefont
  {Crivellin}}, \bibinfo {author} {\bibfnamefont {A.}~\bibnamefont {Kokulu}}, \
  and\ \bibinfo {author} {\bibfnamefont {C.}~\bibnamefont {Greub}},\ }\href
  {\doibase 10.1103/PhysRevD.87.094031} {\bibfield  {journal} {\bibinfo
  {journal} {Phys. Rev. D}\ }\textbf {\bibinfo {volume} {87}},\ \bibinfo
  {pages} {094031} (\bibinfo {year} {2013})},\ \Eprint
  {http://arxiv.org/abs/1303.5877} {arXiv:1303.5877 [hep-ph]} \BibitemShut
  {NoStop}%
\bibitem [{\citenamefont {Kopp}\ and\ \citenamefont
  {Nardecchia}(2014)}]{Kopp:2014rva}%
  \BibitemOpen
  \bibfield  {author} {\bibinfo {author} {\bibfnamefont {J.}~\bibnamefont
  {Kopp}}\ and\ \bibinfo {author} {\bibfnamefont {M.}~\bibnamefont
  {Nardecchia}},\ }\href {\doibase 10.1007/JHEP10(2014)156} {\bibfield
  {journal} {\bibinfo  {journal} {JHEP}\ }\textbf {\bibinfo {volume} {10}},\
  \bibinfo {pages} {156} (\bibinfo {year} {2014})},\ \Eprint
  {http://arxiv.org/abs/1406.5303} {arXiv:1406.5303 [hep-ph]} \BibitemShut
  {NoStop}%
\bibitem [{\citenamefont {Buschmann}\ \emph {et~al.}(2016)\citenamefont
  {Buschmann}, \citenamefont {Kopp}, \citenamefont {Liu},\ and\ \citenamefont
  {Wang}}]{Buschmann:2016uzg}%
  \BibitemOpen
  \bibfield  {author} {\bibinfo {author} {\bibfnamefont {M.}~\bibnamefont
  {Buschmann}}, \bibinfo {author} {\bibfnamefont {J.}~\bibnamefont {Kopp}},
  \bibinfo {author} {\bibfnamefont {J.}~\bibnamefont {Liu}}, \ and\ \bibinfo
  {author} {\bibfnamefont {X.-P.}\ \bibnamefont {Wang}},\ }\href {\doibase
  10.1007/JHEP06(2016)149} {\bibfield  {journal} {\bibinfo  {journal} {JHEP}\
  }\textbf {\bibinfo {volume} {06}},\ \bibinfo {pages} {149} (\bibinfo {year}
  {2016})},\ \Eprint {http://arxiv.org/abs/1601.02616} {arXiv:1601.02616
  [hep-ph]} \BibitemShut {NoStop}%
\bibitem [{\citenamefont {Primulando}\ and\ \citenamefont
  {Uttayarat}(2017)}]{Primulando:2016eod}%
  \BibitemOpen
  \bibfield  {author} {\bibinfo {author} {\bibfnamefont {R.}~\bibnamefont
  {Primulando}}\ and\ \bibinfo {author} {\bibfnamefont {P.}~\bibnamefont
  {Uttayarat}},\ }\href {\doibase 10.1007/JHEP05(2017)055} {\bibfield
  {journal} {\bibinfo  {journal} {JHEP}\ }\textbf {\bibinfo {volume} {05}},\
  \bibinfo {pages} {055} (\bibinfo {year} {2017})},\ \Eprint
  {http://arxiv.org/abs/1612.01644} {arXiv:1612.01644 [hep-ph]} \BibitemShut
  {NoStop}%
\bibitem [{\citenamefont {Altmannshofer}\ \emph {et~al.}(2016)\citenamefont
  {Altmannshofer}, \citenamefont {Eby}, \citenamefont {Gori}, \citenamefont
  {Lotito}, \citenamefont {Martone},\ and\ \citenamefont
  {Tuckler}}]{Altmannshofer:2016zrn}%
  \BibitemOpen
  \bibfield  {author} {\bibinfo {author} {\bibfnamefont {W.}~\bibnamefont
  {Altmannshofer}}, \bibinfo {author} {\bibfnamefont {J.}~\bibnamefont {Eby}},
  \bibinfo {author} {\bibfnamefont {S.}~\bibnamefont {Gori}}, \bibinfo {author}
  {\bibfnamefont {M.}~\bibnamefont {Lotito}}, \bibinfo {author} {\bibfnamefont
  {M.}~\bibnamefont {Martone}}, \ and\ \bibinfo {author} {\bibfnamefont
  {D.}~\bibnamefont {Tuckler}},\ }\href {\doibase 10.1103/PhysRevD.94.115032}
  {\bibfield  {journal} {\bibinfo  {journal} {Phys. Rev. D}\ }\textbf {\bibinfo
  {volume} {94}},\ \bibinfo {pages} {115032} (\bibinfo {year} {2016})},\
  \Eprint {http://arxiv.org/abs/1610.02398} {arXiv:1610.02398 [hep-ph]}
  \BibitemShut {NoStop}%
\bibitem [{\citenamefont {Primulando}\ \emph {et~al.}(2020)\citenamefont
  {Primulando}, \citenamefont {Julio},\ and\ \citenamefont
  {Uttayarat}}]{Primulando:2019ydt}%
  \BibitemOpen
  \bibfield  {author} {\bibinfo {author} {\bibfnamefont {R.}~\bibnamefont
  {Primulando}}, \bibinfo {author} {\bibfnamefont {J.}~\bibnamefont {Julio}}, \
  and\ \bibinfo {author} {\bibfnamefont {P.}~\bibnamefont {Uttayarat}},\ }\href
  {\doibase 10.1103/PhysRevD.101.055021} {\bibfield  {journal} {\bibinfo
  {journal} {Phys. Rev. D}\ }\textbf {\bibinfo {volume} {101}},\ \bibinfo
  {pages} {055021} (\bibinfo {year} {2020})},\ \Eprint
  {http://arxiv.org/abs/1912.08533} {arXiv:1912.08533 [hep-ph]} \BibitemShut
  {NoStop}%
\bibitem [{\citenamefont {Barman}\ \emph {et~al.}(2023)\citenamefont {Barman},
  \citenamefont {Dev},\ and\ \citenamefont {Thapa}}]{Barman:2022iwj}%
  \BibitemOpen
  \bibfield  {author} {\bibinfo {author} {\bibfnamefont {R.~K.}\ \bibnamefont
  {Barman}}, \bibinfo {author} {\bibfnamefont {P.~S.~B.}\ \bibnamefont {Dev}},
  \ and\ \bibinfo {author} {\bibfnamefont {A.}~\bibnamefont {Thapa}},\ }\href
  {\doibase 10.1103/PhysRevD.107.075018} {\bibfield  {journal} {\bibinfo
  {journal} {Phys. Rev. D}\ }\textbf {\bibinfo {volume} {107}},\ \bibinfo
  {pages} {075018} (\bibinfo {year} {2023})},\ \Eprint
  {http://arxiv.org/abs/2210.16287} {arXiv:2210.16287 [hep-ph]} \BibitemShut
  {NoStop}%
\bibitem [{\citenamefont {Hayrapetyan}\ \emph {et~al.}(2023)\citenamefont
  {Hayrapetyan} \emph {et~al.}}]{CMS:2023pte}%
  \BibitemOpen
  \bibfield  {author} {\bibinfo {author} {\bibfnamefont {A.}~\bibnamefont
  {Hayrapetyan}} \emph {et~al.} (\bibinfo {collaboration} {CMS}),\ }\href@noop
  {} {\  (\bibinfo {year} {2023})},\ \Eprint {http://arxiv.org/abs/2305.18106}
  {arXiv:2305.18106 [hep-ex]} \BibitemShut {NoStop}%
\bibitem [{\citenamefont {Georgi}\ and\ \citenamefont
  {Nanopoulos}(1979)}]{Georgi:1978ri}%
  \BibitemOpen
  \bibfield  {author} {\bibinfo {author} {\bibfnamefont {H.}~\bibnamefont
  {Georgi}}\ and\ \bibinfo {author} {\bibfnamefont {D.~V.}\ \bibnamefont
  {Nanopoulos}},\ }\href {\doibase 10.1016/0370-2693(79)90433-7} {\bibfield
  {journal} {\bibinfo  {journal} {Phys. Lett. B}\ }\textbf {\bibinfo {volume}
  {82}},\ \bibinfo {pages} {95} (\bibinfo {year} {1979})}\BibitemShut {NoStop}%
\bibitem [{\citenamefont {Aad}\ \emph {et~al.}(2016{\natexlab{b}})\citenamefont
  {Aad} \emph {et~al.}}]{Khachatryan:2016vau}%
  \BibitemOpen
  \bibfield  {author} {\bibinfo {author} {\bibfnamefont {G.}~\bibnamefont
  {Aad}} \emph {et~al.} (\bibinfo {collaboration} {ATLAS, CMS}),\ }\href
  {\doibase 10.1007/JHEP08(2016)045} {\bibfield  {journal} {\bibinfo  {journal}
  {JHEP}\ }\textbf {\bibinfo {volume} {08}},\ \bibinfo {pages} {045} (\bibinfo
  {year} {2016}{\natexlab{b}})},\ \Eprint {http://arxiv.org/abs/1606.02266}
  {arXiv:1606.02266 [hep-ex]} \BibitemShut {NoStop}%
\bibitem [{\citenamefont {Sirunyan}\ \emph {et~al.}(2019)\citenamefont
  {Sirunyan} \emph {et~al.}}]{CMS:2018uag}%
  \BibitemOpen
  \bibfield  {author} {\bibinfo {author} {\bibfnamefont {A.~M.}\ \bibnamefont
  {Sirunyan}} \emph {et~al.} (\bibinfo {collaboration} {CMS}),\ }\href
  {\doibase 10.1140/epjc/s10052-019-6909-y} {\bibfield  {journal} {\bibinfo
  {journal} {Eur. Phys. J. C}\ }\textbf {\bibinfo {volume} {79}},\ \bibinfo
  {pages} {421} (\bibinfo {year} {2019})},\ \Eprint
  {http://arxiv.org/abs/1809.10733} {arXiv:1809.10733 [hep-ex]} \BibitemShut
  {NoStop}%
\bibitem [{\citenamefont {Aad}\ \emph {et~al.}(2020{\natexlab{b}})\citenamefont
  {Aad} \emph {et~al.}}]{ATLAS:2019nkf}%
  \BibitemOpen
  \bibfield  {author} {\bibinfo {author} {\bibfnamefont {G.}~\bibnamefont
  {Aad}} \emph {et~al.} (\bibinfo {collaboration} {ATLAS}),\ }\href {\doibase
  10.1103/PhysRevD.101.012002} {\bibfield  {journal} {\bibinfo  {journal}
  {Phys. Rev. D}\ }\textbf {\bibinfo {volume} {101}},\ \bibinfo {pages}
  {012002} (\bibinfo {year} {2020}{\natexlab{b}})},\ \Eprint
  {http://arxiv.org/abs/1909.02845} {arXiv:1909.02845 [hep-ex]} \BibitemShut
  {NoStop}%
\bibitem [{\citenamefont {Andreev}\ \emph {et~al.}(2018)\citenamefont {Andreev}
  \emph {et~al.}}]{ACME:2018yjb}%
  \BibitemOpen
  \bibfield  {author} {\bibinfo {author} {\bibfnamefont {V.}~\bibnamefont
  {Andreev}} \emph {et~al.} (\bibinfo {collaboration} {ACME}),\ }\href
  {\doibase 10.1038/s41586-018-0599-8} {\bibfield  {journal} {\bibinfo
  {journal} {Nature}\ }\textbf {\bibinfo {volume} {562}},\ \bibinfo {pages}
  {355} (\bibinfo {year} {2018})}\BibitemShut {NoStop}%
\bibitem [{\citenamefont {de~Florian}\ \emph {et~al.}(2016)\citenamefont
  {de~Florian} \emph {et~al.}}]{LHCHiggsCrossSectionWorkingGroup:2016ypw}%
  \BibitemOpen
  \bibfield  {author} {\bibinfo {author} {\bibfnamefont {D.}~\bibnamefont
  {de~Florian}} \emph {et~al.} (\bibinfo {collaboration} {LHC Higgs Cross
  Section Working Group}),\ }\href {\doibase 10.23731/CYRM-2017-002} {\ \textbf
  {\bibinfo {volume} {2/2017}} (\bibinfo {year} {2016}),\
  10.23731/CYRM-2017-002},\ \Eprint {http://arxiv.org/abs/1610.07922}
  {arXiv:1610.07922 [hep-ph]} \BibitemShut {NoStop}%
\bibitem [{\citenamefont {Chang}\ \emph {et~al.}(1993)\citenamefont {Chang},
  \citenamefont {Hou},\ and\ \citenamefont {Keung}}]{Chang:1993kw}%
  \BibitemOpen
  \bibfield  {author} {\bibinfo {author} {\bibfnamefont {D.}~\bibnamefont
  {Chang}}, \bibinfo {author} {\bibfnamefont {W.~S.}\ \bibnamefont {Hou}}, \
  and\ \bibinfo {author} {\bibfnamefont {W.-Y.}\ \bibnamefont {Keung}},\ }\href
  {\doibase 10.1103/PhysRevD.48.217} {\bibfield  {journal} {\bibinfo  {journal}
  {Phys. Rev. D}\ }\textbf {\bibinfo {volume} {48}},\ \bibinfo {pages} {217}
  (\bibinfo {year} {1993})},\ \Eprint {http://arxiv.org/abs/hep-ph/9302267}
  {arXiv:hep-ph/9302267} \BibitemShut {NoStop}%
\bibitem [{\citenamefont {Baldini}\ \emph {et~al.}(2016)\citenamefont {Baldini}
  \emph {et~al.}}]{MEG:2016leq}%
  \BibitemOpen
  \bibfield  {author} {\bibinfo {author} {\bibfnamefont {A.~M.}\ \bibnamefont
  {Baldini}} \emph {et~al.} (\bibinfo {collaboration} {MEG}),\ }\href {\doibase
  10.1140/epjc/s10052-016-4271-x} {\bibfield  {journal} {\bibinfo  {journal}
  {Eur. Phys. J. C}\ }\textbf {\bibinfo {volume} {76}},\ \bibinfo {pages} {434}
  (\bibinfo {year} {2016})},\ \Eprint {http://arxiv.org/abs/1605.05081}
  {arXiv:1605.05081 [hep-ex]} \BibitemShut {NoStop}%
\bibitem [{\citenamefont {Baldini}\ \emph {et~al.}(2018)\citenamefont {Baldini}
  \emph {et~al.}}]{MEGII:2018kmf}%
  \BibitemOpen
  \bibfield  {author} {\bibinfo {author} {\bibfnamefont {A.~M.}\ \bibnamefont
  {Baldini}} \emph {et~al.} (\bibinfo {collaboration} {MEG II}),\ }\href
  {\doibase 10.1140/epjc/s10052-018-5845-6} {\bibfield  {journal} {\bibinfo
  {journal} {Eur. Phys. J. C}\ }\textbf {\bibinfo {volume} {78}},\ \bibinfo
  {pages} {380} (\bibinfo {year} {2018})},\ \Eprint
  {http://arxiv.org/abs/1801.04688} {arXiv:1801.04688 [physics.ins-det]}
  \BibitemShut {NoStop}%
\bibitem [{\citenamefont {Kitano}\ \emph {et~al.}(2002)\citenamefont {Kitano},
  \citenamefont {Koike},\ and\ \citenamefont {Okada}}]{Kitano:2002mt}%
  \BibitemOpen
  \bibfield  {author} {\bibinfo {author} {\bibfnamefont {R.}~\bibnamefont
  {Kitano}}, \bibinfo {author} {\bibfnamefont {M.}~\bibnamefont {Koike}}, \
  and\ \bibinfo {author} {\bibfnamefont {Y.}~\bibnamefont {Okada}},\ }\href
  {\doibase 10.1103/PhysRevD.76.059902} {\bibfield  {journal} {\bibinfo
  {journal} {Phys. Rev. D}\ }\textbf {\bibinfo {volume} {66}},\ \bibinfo
  {pages} {096002} (\bibinfo {year} {2002})},\ \bibinfo {note} {[Erratum:
  Phys.Rev.D 76, 059902 (2007)]},\ \Eprint
  {http://arxiv.org/abs/hep-ph/0203110} {arXiv:hep-ph/0203110} \BibitemShut
  {NoStop}%
\bibitem [{\citenamefont {Crivellin}\ \emph {et~al.}(2014)\citenamefont
  {Crivellin}, \citenamefont {Hoferichter},\ and\ \citenamefont
  {Procura}}]{Crivellin:2014cta}%
  \BibitemOpen
  \bibfield  {author} {\bibinfo {author} {\bibfnamefont {A.}~\bibnamefont
  {Crivellin}}, \bibinfo {author} {\bibfnamefont {M.}~\bibnamefont
  {Hoferichter}}, \ and\ \bibinfo {author} {\bibfnamefont {M.}~\bibnamefont
  {Procura}},\ }\href {\doibase 10.1103/PhysRevD.89.093024} {\bibfield
  {journal} {\bibinfo  {journal} {Phys. Rev. D}\ }\textbf {\bibinfo {volume}
  {89}},\ \bibinfo {pages} {093024} (\bibinfo {year} {2014})},\ \Eprint
  {http://arxiv.org/abs/1404.7134} {arXiv:1404.7134 [hep-ph]} \BibitemShut
  {NoStop}%
\bibitem [{\citenamefont {Bishara}\ \emph {et~al.}(2016)\citenamefont
  {Bishara}, \citenamefont {Brod}, \citenamefont {Uttayarat},\ and\
  \citenamefont {Zupan}}]{Bishara:2015cha}%
  \BibitemOpen
  \bibfield  {author} {\bibinfo {author} {\bibfnamefont {F.}~\bibnamefont
  {Bishara}}, \bibinfo {author} {\bibfnamefont {J.}~\bibnamefont {Brod}},
  \bibinfo {author} {\bibfnamefont {P.}~\bibnamefont {Uttayarat}}, \ and\
  \bibinfo {author} {\bibfnamefont {J.}~\bibnamefont {Zupan}},\ }\href
  {\doibase 10.1007/JHEP01(2016)010} {\bibfield  {journal} {\bibinfo  {journal}
  {JHEP}\ }\textbf {\bibinfo {volume} {01}},\ \bibinfo {pages} {010} (\bibinfo
  {year} {2016})},\ \Eprint {http://arxiv.org/abs/1504.04022} {arXiv:1504.04022
  [hep-ph]} \BibitemShut {NoStop}%
\bibitem [{\citenamefont {Bertl}\ \emph {et~al.}(2006)\citenamefont {Bertl}
  \emph {et~al.}}]{Bertl:2006up}%
  \BibitemOpen
  \bibfield  {author} {\bibinfo {author} {\bibfnamefont {W.~H.}\ \bibnamefont
  {Bertl}} \emph {et~al.} (\bibinfo {collaboration} {SINDRUM II}),\ }\href
  {\doibase 10.1140/epjc/s2006-02582-x} {\bibfield  {journal} {\bibinfo
  {journal} {Eur. Phys. J. C}\ }\textbf {\bibinfo {volume} {47}},\ \bibinfo
  {pages} {337} (\bibinfo {year} {2006})}\BibitemShut {NoStop}%
\bibitem [{\citenamefont {{CMS Collaboration}}(2023)}]{CMS:2023pqk}%
  \BibitemOpen
  \bibfield  {author} {\bibinfo {author} {\bibnamefont {{CMS Collaboration}}},\
  }\href@noop {} {\  (\bibinfo {year} {2023})},\ \Eprint
  {http://arxiv.org/abs/CMS-PAS-HIG-22-022} {CMS-PAS-HIG-22-022} \BibitemShut
  {NoStop}%
\bibitem [{\citenamefont {Cowan}\ \emph {et~al.}(2011)\citenamefont {Cowan},
  \citenamefont {Cranmer}, \citenamefont {Gross},\ and\ \citenamefont
  {Vitells}}]{Cowan_2011}%
  \BibitemOpen
  \bibfield  {author} {\bibinfo {author} {\bibfnamefont {G.}~\bibnamefont
  {Cowan}}, \bibinfo {author} {\bibfnamefont {K.}~\bibnamefont {Cranmer}},
  \bibinfo {author} {\bibfnamefont {E.}~\bibnamefont {Gross}}, \ and\ \bibinfo
  {author} {\bibfnamefont {O.}~\bibnamefont {Vitells}},\ }\href {\doibase
  10.1140/epjc/s10052-011-1554-0} {\bibfield  {journal} {\bibinfo  {journal}
  {The European Physical Journal C}\ }\textbf {\bibinfo {volume} {71}}
  (\bibinfo {year} {2011}),\ 10.1140/epjc/s10052-011-1554-0}\BibitemShut
  {NoStop}%
\bibitem [{\citenamefont {Alwall}\ \emph {et~al.}(2014)\citenamefont {Alwall},
  \citenamefont {Frederix}, \citenamefont {Frixione}, \citenamefont {Hirschi},
  \citenamefont {Maltoni}, \citenamefont {Mattelaer}, \citenamefont {Shao},
  \citenamefont {Stelzer}, \citenamefont {Torrielli},\ and\ \citenamefont
  {Zaro}}]{Alwall:2014hca}%
  \BibitemOpen
  \bibfield  {author} {\bibinfo {author} {\bibfnamefont {J.}~\bibnamefont
  {Alwall}}, \bibinfo {author} {\bibfnamefont {R.}~\bibnamefont {Frederix}},
  \bibinfo {author} {\bibfnamefont {S.}~\bibnamefont {Frixione}}, \bibinfo
  {author} {\bibfnamefont {V.}~\bibnamefont {Hirschi}}, \bibinfo {author}
  {\bibfnamefont {F.}~\bibnamefont {Maltoni}}, \bibinfo {author} {\bibfnamefont
  {O.}~\bibnamefont {Mattelaer}}, \bibinfo {author} {\bibfnamefont {H.~S.}\
  \bibnamefont {Shao}}, \bibinfo {author} {\bibfnamefont {T.}~\bibnamefont
  {Stelzer}}, \bibinfo {author} {\bibfnamefont {P.}~\bibnamefont {Torrielli}},
  \ and\ \bibinfo {author} {\bibfnamefont {M.}~\bibnamefont {Zaro}},\ }\href
  {\doibase 10.1007/JHEP07(2014)079} {\bibfield  {journal} {\bibinfo  {journal}
  {JHEP}\ }\textbf {\bibinfo {volume} {07}},\ \bibinfo {pages} {079} (\bibinfo
  {year} {2014})},\ \Eprint {http://arxiv.org/abs/1405.0301} {arXiv:1405.0301
  [hep-ph]} \BibitemShut {NoStop}%
\bibitem [{\citenamefont {Bierlich}\ \emph {et~al.}(2022)\citenamefont
  {Bierlich} \emph {et~al.}}]{Bierlich:2022pfr}%
  \BibitemOpen
  \bibfield  {author} {\bibinfo {author} {\bibfnamefont {C.}~\bibnamefont
  {Bierlich}} \emph {et~al.},\ }\href {\doibase 10.21468/SciPostPhysCodeb.8} {\
   (\bibinfo {year} {2022}),\ 10.21468/SciPostPhysCodeb.8},\ \Eprint
  {http://arxiv.org/abs/2203.11601} {arXiv:2203.11601 [hep-ph]} \BibitemShut
  {NoStop}%
\bibitem [{\citenamefont {de~Favereau}\ \emph {et~al.}(2014)\citenamefont
  {de~Favereau}, \citenamefont {Delaere}, \citenamefont {Demin}, \citenamefont
  {Giammanco}, \citenamefont {Lema\^\i{}tre}, \citenamefont {Mertens},\ and\
  \citenamefont {Selvaggi}}]{deFavereau:2013fsa}%
  \BibitemOpen
  \bibfield  {author} {\bibinfo {author} {\bibfnamefont {J.}~\bibnamefont
  {de~Favereau}}, \bibinfo {author} {\bibfnamefont {C.}~\bibnamefont
  {Delaere}}, \bibinfo {author} {\bibfnamefont {P.}~\bibnamefont {Demin}},
  \bibinfo {author} {\bibfnamefont {A.}~\bibnamefont {Giammanco}}, \bibinfo
  {author} {\bibfnamefont {V.}~\bibnamefont {Lema\^\i{}tre}}, \bibinfo {author}
  {\bibfnamefont {A.}~\bibnamefont {Mertens}}, \ and\ \bibinfo {author}
  {\bibfnamefont {M.}~\bibnamefont {Selvaggi}} (\bibinfo {collaboration}
  {DELPHES 3}),\ }\href {\doibase 10.1007/JHEP02(2014)057} {\bibfield
  {journal} {\bibinfo  {journal} {JHEP}\ }\textbf {\bibinfo {volume} {02}},\
  \bibinfo {pages} {057} (\bibinfo {year} {2014})},\ \Eprint
  {http://arxiv.org/abs/1307.6346} {arXiv:1307.6346 [hep-ex]} \BibitemShut
  {NoStop}%
\bibitem [{\citenamefont {Sirunyan}\ \emph
  {et~al.}(2018{\natexlab{b}})\citenamefont {Sirunyan} \emph
  {et~al.}}]{CMS:2018amk}%
  \BibitemOpen
  \bibfield  {author} {\bibinfo {author} {\bibfnamefont {A.~M.}\ \bibnamefont
  {Sirunyan}} \emph {et~al.} (\bibinfo {collaboration} {CMS}),\ }\href
  {\doibase 10.1007/JHEP06(2018)127} {\bibfield  {journal} {\bibinfo  {journal}
  {JHEP}\ }\textbf {\bibinfo {volume} {06}},\ \bibinfo {pages} {127} (\bibinfo
  {year} {2018}{\natexlab{b}})},\ \bibinfo {note} {[Erratum: JHEP 03, 128
  (2019)]},\ \Eprint {http://arxiv.org/abs/1804.01939} {arXiv:1804.01939
  [hep-ex]} \BibitemShut {NoStop}%
\bibitem [{\citenamefont {{CMS
  Collaboration}}(2022{\natexlab{c}})}]{CMS:2022goy}%
  \BibitemOpen
  \bibfield  {author} {\bibinfo {author} {\bibnamefont {{CMS Collaboration}}},\
  }\href@noop {} {\  (\bibinfo {year} {2022}{\natexlab{c}})},\ \Eprint
  {http://arxiv.org/abs/2208.02717} {arXiv:2208.02717 [hep-ex]} \BibitemShut
  {NoStop}%
\bibitem [{\citenamefont {Tumasyan}\ \emph
  {et~al.}(2022{\natexlab{b}})\citenamefont {Tumasyan} \emph
  {et~al.}}]{CMS:2021cox}%
  \BibitemOpen
  \bibfield  {author} {\bibinfo {author} {\bibfnamefont {A.}~\bibnamefont
  {Tumasyan}} \emph {et~al.} (\bibinfo {collaboration} {CMS}),\ }\href
  {\doibase 10.1007/JHEP04(2022)147} {\bibfield  {journal} {\bibinfo  {journal}
  {JHEP}\ }\textbf {\bibinfo {volume} {04}},\ \bibinfo {pages} {147} (\bibinfo
  {year} {2022}{\natexlab{b}})},\ \Eprint {http://arxiv.org/abs/2106.14246}
  {arXiv:2106.14246 [hep-ex]} \BibitemShut {NoStop}%
\bibitem [{\citenamefont {Crivellin}\ \emph {et~al.}(2021)\citenamefont
  {Crivellin}, \citenamefont {Fang}, \citenamefont {Fischer}, \citenamefont
  {Kumar}, \citenamefont {Kumar}, \citenamefont {Malwa}, \citenamefont
  {Mellado}, \citenamefont {Rapheeha}, \citenamefont {Ruan},\ and\
  \citenamefont {Sha}}]{Crivellin:2021ubm}%
  \BibitemOpen
  \bibfield  {author} {\bibinfo {author} {\bibfnamefont {A.}~\bibnamefont
  {Crivellin}}, \bibinfo {author} {\bibfnamefont {Y.}~\bibnamefont {Fang}},
  \bibinfo {author} {\bibfnamefont {O.}~\bibnamefont {Fischer}}, \bibinfo
  {author} {\bibfnamefont {A.}~\bibnamefont {Kumar}}, \bibinfo {author}
  {\bibfnamefont {M.}~\bibnamefont {Kumar}}, \bibinfo {author} {\bibfnamefont
  {E.}~\bibnamefont {Malwa}}, \bibinfo {author} {\bibfnamefont
  {B.}~\bibnamefont {Mellado}}, \bibinfo {author} {\bibfnamefont
  {N.}~\bibnamefont {Rapheeha}}, \bibinfo {author} {\bibfnamefont
  {X.}~\bibnamefont {Ruan}}, \ and\ \bibinfo {author} {\bibfnamefont
  {Q.}~\bibnamefont {Sha}},\ }\href@noop {} {\  (\bibinfo {year} {2021})},\
  \Eprint {http://arxiv.org/abs/2109.02650} {arXiv:2109.02650 [hep-ph]}
  \BibitemShut {NoStop}%
\end{thebibliography}%
\bibliographystyle{apsrev4-1}

\end{document}